# Reply to "Extracting Kondo temperature of strongly-correlated systems from the inverse local magnetic susceptibility" by A. A. Katanin

Xiaoyu Deng[1], Katharina M. Stadler[2], Kristjan Haule[1], Seung-Sup B. Lee[2], Andreas Weichselbaum[2,3], Jan von Delft[2], and Gabriel Kotliar[1,3]

1: Department of Physics and Astronomy, Rutgers University, Piscataway, NJ, USA

2: Physics Department, Arnold Sommerfeld Center for Theoretical Physics and Center for NanoScience, Ludwig-Maximilians-Universitat München, München, Germany

3: Condensed Matter Physics and Materials Science Department, Brookhaven National Laboratory, Upton, NY, USA

In his comment [1], Katanin reanalyzes our LDA+DMFT results [2] for the temperature-dependent static local spin susceptibility of $Sr_2RuO_4$ and $V_2O_3$, fitting them to a Curie-Weiss (CW) form, $\chi(T) \simeq a/(T+\theta)$. Invoking Wilson's analysis [3] of the impurity susceptibility of the spin-½ one-channel Kondo model (1CKM) in the wide-band limit, he extracts spin Kondo temperatures using $T_K = \theta/\sqrt{2}$, obtaining $T_K$ = 350 K and 100 K for $Sr_2RuO_4$ and $V_2O_3$, respectively. Noting that these are significantly smaller than the scales $T_{sp}^{onset}$= 2300 K and 1000K reported in Ref. 2, he argues that our $T_{sp}^{onset}$ scales "do not characterize the screening process".

We welcome Katanin's use of our data. However, his implication that our $T_{sp}^{onset}$ was intended to fully characterize the screening process is misleading. Our work uses the full susceptibility vs. temperature curve to describe properly spin screening, not just a single number. Furthermore our $T_{sp}^{onset}$ was defined to characterize the high-temperature *onset* of spin screening, whereas his $T_K$ characterizes the CW regime found at intermediate (i.e., lower) temperatures. The fact that $T_K$ is much smaller than $T_{sp}^{onset}$ is therefore not surprising but natural.

We agree with Katanin that, for Hund metals in general and $Sr_2RuO_4$ in particular, it is reasonable to approximate $\chi(T)$ using results of a Kondo impurity which feature a CW law at intermediate temperatures. (In the Supplementary material we analyze $\chi(T)$ taken data from DMFT studies of the model Hund system used in Ref. 2.) However, this was already well known. For $Sr_2RuO_4$, a comparison to the exact solution of a (fully screened) spin-1 Kondo impurity model was carried out in the inset of Fig. 3(a) of Ref. 4 (Ref. 17 of Ref. 2), reproduced as Fig. 1(left) below, and a CW fit of that data was published in Fig. 2(a) of Ref. 5 (cited as Ref. 5 of Ref. 2). We reproduce them as Fig. 1(right) below. Since $Sr_2RuO_4$ and $V_2O_3$ have an atomic ground state configuration spin closer to 1 than ½, the use of a (fully-screened) spin-1 Kondo model is more reasonable. Furthermore, when interpreting LDA+DMFT results, it is preferable to use definitions of the Kondo scale that rely on the *low*-temperature portion of the susceptibility curve, as was done in Refs. 4 and 10, as opposed to the high-temperature as in Katanin's proposal to characterize spin screening. We elaborate on these points and propose a simple way to characterize spin crossovers of Hund metals below.

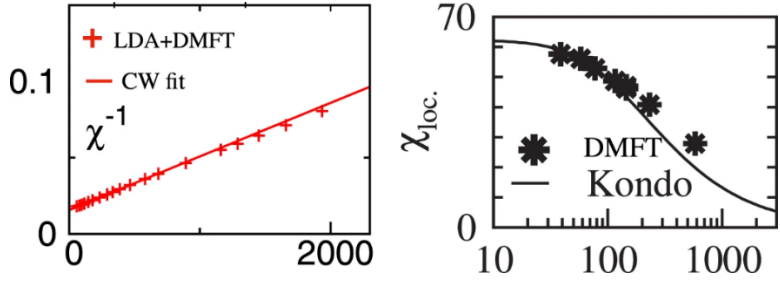

**Fig. 1: Earlier work comparing Kondo impurity model with LDA+DMFT results for $Sr_2RuO_4$.** $1/\chi(T)$ versus $T$, with LDA+DMFT results for $Sr_2RuO_4$ (red symbols) and a Curie–Weiss fit (straight red line) reproduced from the inset of Fig. 2(a) of Ref. 5. Right: Bethe-Ansatz results for the spin-1,2-channel Kondo model $\chi(T)$ vs. T with $T_{BA}$ = 240 K (solid line) in good agreement with the LDA + DMFT results for $Sr_2RuO_4$ (black symbols) reproduced from inset of Fig. 3(a) of Ref. 4.

Since Katanin's comment invokes the 1CKM, we start by summarizing some of its well-established properties [3, 6, 7]. $\chi(T)$ exhibits a very broad crossover, from Curie-like high-temperature behavior governed by a local-moment fixed point describing a free spin, to Pauli-like low-temperature behavior governed by a Fermi-liquid fixed point describing a fully screened spin. A proper description of this crossover requires a crossover scaling function, $F(T/T_K)$ and a crossover scale, the Kondo scale $T_K$, with $\chi(T) = F(T/T_K)/T$. Wilson showed that $F(x)$ is universal under the assumptions of very weak impurity-bath coupling and infinite bandwidth, and computed it numerically. There are multiple ways of defining $T_K$, evoking the behavior of $F(x)$ for either $x \gg 1$, $x \simeq 1$ or $x \ll 1$, yielding $T_K$ values differing only by factors of order unity. Wilson's definition of $T_K$ (adopted by Katanin), denoted $T_W$ here, evokes the $x \gg 1$ limit. For high temperatures, $T \gtrsim 16 T_W$, he found $\chi(T) \simeq 1/(4T) [1 - 1/\ln(T/T_W) + O(1/\ln^3(T/T_W)]$, with $T_W$ defined such that the coefficient of $1/\ln^2(T/T_W)$ vanishes. For intermediate temperatures, $0.5\,T_W < T < 16\,T_W$, his numerical results are well approximated by a CW form, with $a = 0.17$ and $\theta \sim \sqrt{2}\,T_W$ [3, 6] (as used by Katanin). At zero temperature, Wilson found $\chi(0) \sim 0.103\,/T_W$ (Eq. (IX.91) of Ref. 3). Subsequent Bethe-Ansatz (BA) calculations of the scaling function [6, 7] matched Wilson's numerical results. Analogous results have been obtained for fully-screened Kondo models with higher spins [8, 9]. The BA works showed that the curve $\chi(T)$ vs. $T/T_K$ depends on the spin $S$, with $\chi(T) \propto S(S+1)/(3T)$ for $T/T_K \gg 1$ and $\chi(T) \propto S$ for $T/T_K \ll 1$. The Kondo scales defined in these BA works are independent of spin, as in Eq. (21) of Ref. 9: $T_{BA} = S/[\pi\chi(0)]$, with $T_{BA}/T_W = 1.55$ for $S = 1/2$.

In Ref. 2, we used a strategy similar to Wilson's: we identified the regions where the behavior of $\chi_{spin}(T)$ and $\chi_{orb}(T)$ is governed by atomic physics or Fermi-liquid theory and numerically computed the crossover function bridging them. We defined two scales for the onset and completion of spin screening, $T_{sp}^{onset}$ and $T_{sp}^{cmp}$, as the temperatures above or below which $\chi_{spin}(T)$ shows pure Curie behavior ($\sim 1/T$) or pure Pauli behavior ($\sim$ const.), respectively, and similarly $T_{orb}^{onset}$ and $T_{orb}^{cmp}$ for orbital screening. Our $T_{sp}^{onset}$ and $T_{sp}^{cmp}$ scales are similar in spirit to Wilson's $16T_W$ and $0.5T_W$. So even within the 1CKM framework, an extraction of $T_W$ from our results, using $T_W \simeq T_{sp}^{onset}/16$, would yield 2300 K/16 $\simeq$ 140 K for $Sr_2RuO_4$ and 1000 K/16 $\simeq$ 60 K, and the order of magnitude discrepancy claimed by Katanin disappears.

Contrary to this crude estimate, in Ref. 2 we did *not* assume $T_{\text{sp}}^{\text{onset}}$ to be proportional to a single Kondo scale, since even for an impurity model without DMFT self-consistency, $T_{\text{sp}}^{\text{onset}}$ is known to be affected by energy scales not present in the wide-band 1CKM (e.g., a finite bandwidth or a finite charging energy), since such scales cut off high-temperature logarithmic corrections [cf. Ref. 10, Fig. 2(b,c)]. This is even more important for Mott systems, where the emergence of a quasi-particle resonance with decreasing temperatures affects the bath bandwidth via DMFT self-consistency.

In Ref. 2, we supplemented our LDA+DMFT study of actual materials by DMFT studies of a multi-orbital model Hamiltonian, again computing $\chi(T)$ numerically. We found signatures distinguishing Mottness and Hundness (such as $T_{\text{spin}}^{\text{onset}} \simeq T_{\text{orb}}^{\text{onset}}$ for the former but $T_{\text{spin}}^{\text{onset}} < T_{\text{orb}}^{\text{onset}}$ for the latter) similar to those found in the materials. We defined a Kondo scale $T_{\text{K,spin}}^{\text{dyn}}$ (denoted $T_{\text{K}}$ in Ref. 2) through the imaginary part of the $T=0$ dynamical spin susceptibility, $\chi''\left(\omega = T_{\text{K,spin}}^{\text{dyn}}\right) =$ maximal. $T_{\text{K,spin}}^{\text{dyn}}$ characterizes the intermediate region, with $T_{\text{spin}}^{\text{cmp}} < T_{\text{K,spin}}^{\text{dyn}} < T_{\text{spin}}^{\text{onset}}$. It is shown as a red line in Fig. 5b of Ref. 2, yielding $T_{\text{K,spin}}^{\text{dyn}} = 0.12t = 600$ K for our Hund system H1 mimicking Sr$_2$RuO$_4$, and $T_{\text{K,spin}}^{\text{dyn}} = 0.04t = 200$ K for our Mott system M1 mimicking V$_2$O$_3$ (using the conversion factor $t = 5000$ K stated in Fig. 1).

We take Katanin's comment as an incentive to propose a standardized scheme for extracting a Kondo scale, $\tilde{T}_{\text{K}}$, from a computed $\chi(T)$ curve. Our scheme: (i) does not involve a fit to predictions of a specific impurity model, since in general it is unclear which impurity model to compare to; (ii) uses the $x \leq 1$ part of the crossover scaling function, since it is more universal than the $x \gg 1$ part [8, 9, 10]; and (iii) reduces to impurity-model results when these are applicable. We propose to define $\tilde{T}_{\text{K}}$ through the relation $\chi(\tilde{T}_{\text{K}})/\chi(0) = 1/2$. (If $\chi(0)$ is not known but $\chi(T)$ shows CW-type behavior at intermediate temperatures, $\chi(0)$ can be estimated by linear extrapolation of $1/\chi(T)$ vs. $T$ to zero temperature.) This definition ensures that $T_{\text{sp}}^{\text{comp}} < \tilde{T}_{\text{K}} < T_{\text{sp}}^{\text{onset}}$, as it should. For the CW form it yields $\tilde{T}_{\text{K}} = \theta$. For the 1CKM, NRG computations (Fig. 2) show that $\tilde{T}_{\text{K}} = 0.169/\chi(0) = 1.06\, T_{\text{BA}} = 1.64\, T_{\text{W}}$. For the materials Sr$_2$RuO$_4$ and V$_2$O$_3$ studied in Ref. 2, Katanin's CW extraction of $\theta$-values implies $\tilde{T}_{\text{K}} = 574$ K or 164 K, respectively. This illustrates, yet again, the main point of this reply: the Kondo scale is generically much smaller than $T_{\text{sp}}^{\text{onset}}$, and it is misleading to conflate these two scales.

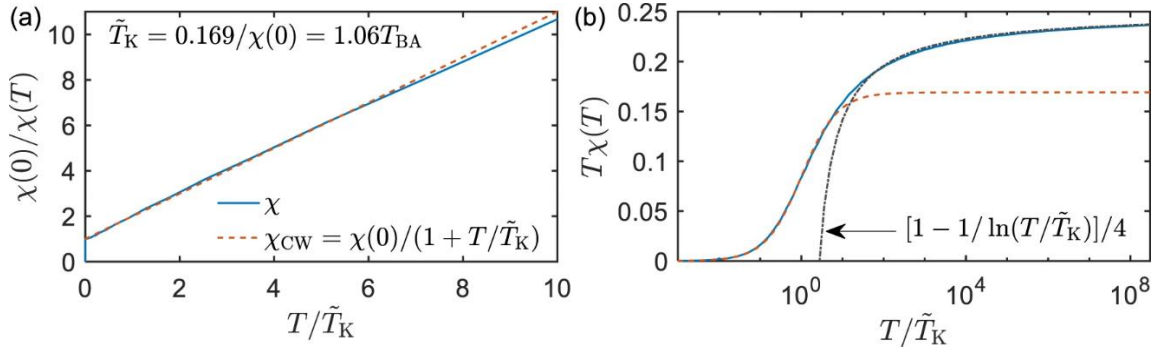

**Fig. 2: Two representations of the impurity susceptibility $\chi(T)$ as defined by Wilson [3] (blue lines), for the 1CKM in the wide-band limit, computed using the numerical renormalization group (NRG).** (a) The Curie–Weiss form (red dashed line) works reasonably well for intermediate temperatures, but (b) not at all for large temperatures, $T/\tilde{T}_{\text{K}} \gg 1$, where logarithmic corrections are large (black dash-dotted line).

**Supplementary material:**

The Curie-Weiss law, $\chi(T) = \mu/(T + \theta)$, is often used to roughly characterize the temperature dependence of the susceptibility of Kondo impurities and of the local susceptibility of Hund metals. Correspondingly, a plot of $1/\chi(T)$ vs. $T$ should give a straight line with slope $1/\mu$ and y-axis intercept $\theta/\mu$. Here we briefly elaborate on the extent to which this is the case for the data which we had published in Ref. 2.

In our publication, we had computed $\chi_0^{sp}(T)$, defined there as the static ($\omega \to 0$ limit of the dynamic) local spin susceptibility, for two setups: (i) realistic LDA+DMFT calculations for the material systems $V_2O_3$ and $Sr_2RuO_4$; and (ii) DMFT+NRG calculations for a Mott system M1 and a Hund system H1 in the context of a multi-orbital Hund-Hubbard model, which offers a minimal description of the physics of Hund's metals, and allows us to extract more precise values of susceptibility over several decades in temperature . In his comment, Katanin plotted $1/\chi_0^{sp}(T)$ vs.$T$ using our data from (i). In panels (a,b) of Fig. S1 below, we show similar plots obtained using our data from (ii). In both Katanin's plots and ours, the resulting curves can be fit fairly well by a "global" straight line $y(T) = (T + \theta_{glo})/\mu_{glo}$, but deviations are clearly discernable.

To quantify the extent to which a global Curie-Weiss description is too simplistic, we perform local analysis, viewing $\mu$ and $\theta$ not as constants, but as $T$-dependent parameters characterizing straight lines tangent to the inverse susceptibility curve. Concretely, for each value of $T$, we define the function $y(T') = [T' + \theta(T)]/\mu(T)$ as the straight line which is tangent to the curve $1/\chi_0^{sp}(T')$ at $T' = T$, i.e., the latter curve has "local" slope $1/\mu(T)$, and its tangent crosses the $y$ axis at $\theta(T)/\mu(T)$. Hence we define the temperature dependent moment and Curie Weiss temperature as $1/\mu(T) = d(1/\chi)/dT$, and $\theta(T)/\mu(T) = 1/\chi - T/\mu(T)$.

Panels (c,d) of Fig. S1 show the functions $\mu(T)$ and $\theta(T)$ so obtained for the Mott system M1 and the Hund system H1. They are not constant, indicating the limitations of a pure Curie-Weiss description.

The slight maximum in $\mu(T)$ for M1 in panel (c), near $T_{max} \simeq 0.1$, reflects a slight decrease in the slope of $1/\chi_0^{sp}(T)$ seen at that temperature in panel (a), which is also mentioned by Katanin in Ref. 1 for the material system $V_2O_3$. Note that $T_{max}$ is about 5 times larger than the Curie-Weiss scale $\theta_{glo}$. It reflects physics beyond that of a simple impurity model, requiring DMFT self-consistency instead, namely $T_{max}$ is the onset of metallicity in the Mott system as the temperature is lowered. Indeed, the value of $T_{max}$ agrees, up to a factor of two, with the temperature $T_M = T_{spin}^{onset} \simeq 0.2$ which we had identified for the Mott system M1 in our paper as the temperature where a coherence resonance emerges from the pseudogap as the temperature is lowered.

Figure S1 thus provides additional evidence supporting our assertion that for Mott systems, the onset of the Kondo resonance is not connected to logarithmic singularities, but instead is driven by the DMFT-self-consistency condition, as is well known in the DMFT literature.

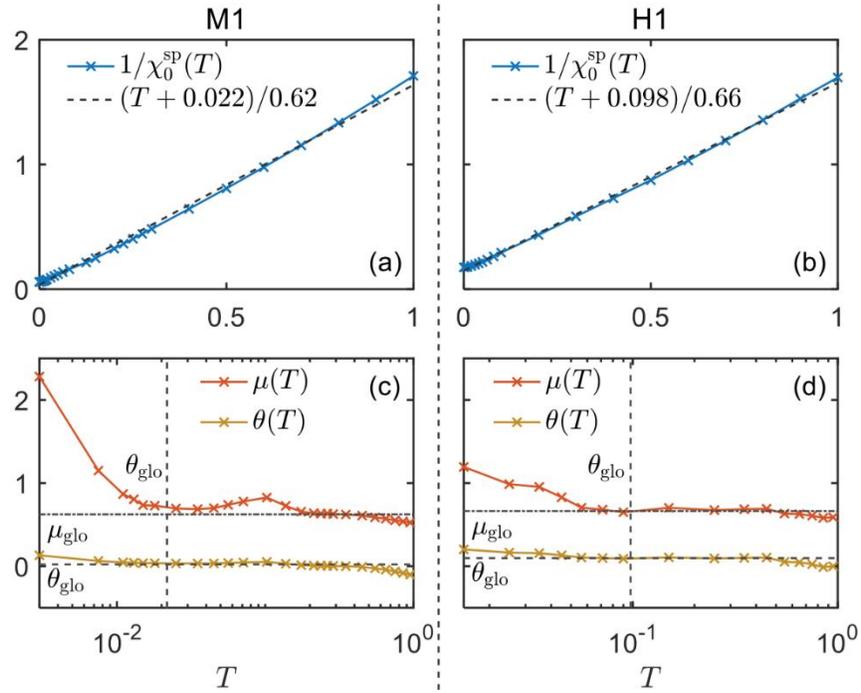

Fig. S1: Curie-Weiss analysis of the temperature dependence of the static local spin susceptibility $\chi_0^{\mathrm{sp}}(T)$, obtained from DMFT+NRG calculations [2] for (a,c) a Mott system (M1; U=6.5, J=1) and (b,d) a Hund system (H1; U=3, J=1). (a,b) Global Curie-Weiss fits. (c,d) The functions $\mu(T)$ and $\theta(T)$ characterizing straight lines that are tangent to the inverse susceptibility (see text). The parameters $\mu_{\mathrm{glo}}$ and $\theta_{\mathrm{glo}}$ obtained from (a,b) are indicated in (c,d) by dashed-dotted or dashed lines, respectively.